\documentclass[twocolumn,prd,10pt,nofootinbib,superscriptaddress]{revtex4-1}
\usepackage{}
\usepackage{epsfig,latexsym,amssymb}
\usepackage{latexsym}
\usepackage{amsmath}
\usepackage{amssymb}
\usepackage{amsfonts}
\usepackage{comment}
\usepackage{graphicx}
\usepackage{verbatim}
\usepackage{epstopdf}
\usepackage{hyperref, url}
\hypersetup{colorlinks   = true,
            urlcolor     = blue,
            citecolor    = blue,
            linkcolor    = blue,
            menucolor    = blue,
            anchorcolor  = blue,
            filecolor    = blue}

\newcommand{\be}{\begin{equation}}
\newcommand{\ee}{\end{equation}}
\newcommand{\bea}{\begin{eqnarray}}
\newcommand{\eea}{\end{eqnarray}}

\begin{document}

\title{Detecting dilute axion stars constrained by fast radio bursts \\in the Solar System via stimulated decay}

\author{Haoran Di}
\email[Corresponding author: ]{hrdi@ecut.edu.cn}
\affiliation{School of Science, East China University of Technology, Nanchang 330013, China}
\author{Zhu Yi}
\affiliation{Faculty of Arts and Sciences, Beijing Normal University, Zhuhai 519087, China}
\affiliation{Advanced Institute of Natural Sciences, Beijing Normal University, Zhuhai 519087, China}
\author{Haihao Shi}
\affiliation{Xinjiang Astronomical Observatory, CAS, Urumqi 830011, China}
\affiliation{College of Astronomy and Space Science, University of Chinese Academy of Sciences, Beijing 101408, China}
\author{Yungui Gong}
\email[Corresponding author: ]{gongyungui@nbu.edu.cn}
\affiliation{Institute of Fundamental Physics and Quantum Technology, Department of Physics, School of Physical Science and Technology, Ningbo University, Ningbo 315211, China}

\begin{abstract}
Fast radio bursts (FRBs) can be explained by collapsing axion stars, imposing constraints on the axion parameter space and providing valuable guidance for experimental axion searches. In the traditional post-inflationary model, axion stars could constitute up to $75\%$ of the dark matter component, suggesting that some axion stars may exist within the Solar System. Photons with energy half the axion mass can stimulate axion decay. Thus, directing a powerful radio beam at an axion star could trigger its stimulated decay, producing a detectable echo. Using this method, we find it is possible to test the existence of dilute axion stars with maximum masses ranging from $6.21\times10^{-12}M_\odot$ to $2.61\times10^{-10}M_\odot$, as constrained by FRBs, within the Solar System. The resulting echo from axion stars constrained by FRBs could be detectable by terrestrial telescopes. Detecting such an echo would confirm the existence of axion stars, unravel the mystery of dark matter, and provide key evidence that some FRBs originate from collapsing axion stars. Furthermore, FRBs produced by axion star collapses could serve as standard candles, aiding in the resolution of the Hubble tension. If no echo is detected using this method, it would place constraints on the abundance of dark matter in the form of dilute axion stars with maximum masses in the range of $6.21\times10^{-12}M_\odot$ to $2.61\times10^{-10}M_\odot$.
\end{abstract}

\maketitle

\section{Introduction}
Growing evidence from various observations and theoretical frameworks has established that dark matter constitutes a substantial portion of the Universe's energy density. Nonetheless, its exact nature remains elusive. The QCD axion \cite{Weinberg:1977ma,Wilczek:1977pj}, arising from the Peccei-Quinn mechanism \cite{Peccei:1977ur,Peccei:1977hh} developed to resolve the strong-$CP$ problem, is a well-motivated candidate for dark matter. Additionally, string theory predicts a spectrum of axion-like particles (ALPs) across a wide mass range, a concept often referred to as the ``axiverse" \cite{Arvanitaki:2009fg}. In this article, we use the term ``axions" to encompass both QCD axions and ALPs.
Axions can be produced through several mechanisms, such as the misalignment mechanism \cite{Preskill:1982cy,Abbott:1982af,Dine:1982ah,Co:2019jts}, thermal production \cite{Turner:1986tb,Salvio:2013iaa}, the decay of string \cite{Davis:1986xc,Gorghetto:2020qws}, and primordial black hole (PBH) evaporation \cite{Schiavone:2021imu,Li:2022xqh,Mazde:2022sdx}. Because axions are bosons, they can attain very high phase space densities, which may result in Bose-Einstein condensation (BEC) \cite{Sikivie:2009qn} and the formation of gravitationally bound objects known as axion stars \cite{Braaten:2019knj,Visinelli:2021uve,Gorghetto:2024vnp,Chang:2024fol,Zhang:2024bjo}.

The unusual orbits of trans-Neptunian objects (TNOs) \cite{Brown:2004yy,Trujillo,Batygin:2016zsa} and the gravitational anomalies detected by the Optical Gravitational Lensing Experiment (OGLE) \cite{Mroz:2017mvf} remain unexplained. One potential explanation is the existence of PBHs \cite{Niikura:2019kqi} or axion stars \cite{Sugiyama:2021xqg} with masses in the range of $M \sim 0.5 - 20M_\oplus$, which could account for the OGLE observations. To address the TNO orbital anomalies, the planet 9 hypothesis proposes an object with a mass of $5-15M_\oplus$ (approximately $1.5-4.5\times10^{-5}M_\odot$) located $300-1000$ AU from the Sun \cite{Batygin-PR}. This hypothetical planet 9 could be explained as a dilute axion star \cite{Di:2023xaw} that was captured by the Solar System.
Additionally, collapsing axion stars with specific parameters may emit millisecond-long radio bursts with peak luminosities of approximately $10^{42}$ erg/s, which are consistent with the characteristics of observed non-repeating fast radio bursts (FRBs) \cite{Di:2023nnb}. These collapsing axion stars could serve as novel standard candles \cite{Di:2024tlz} for constraining the Hubble constant, $H_0$, due to their strong, intrinsic luminosity being fixed and dependent solely on the axion mass and decay constant. Other studies have also investigated the relationship between axion stars and FRBs \cite{Iwazaki:2014wka,Tkachev:2014dpa,Raby:2016deh,Buckley:2020fmh}. If some FRBs are indeed produced by collapsing axion stars, this would impose constraints on the axion parameter space \cite{Di:2024tlz}, providing valuable guidance for experimental searches for axions. Thus, exploring the axion parameter space in relation to FRBs is crucial, as it offers insight into the origin of FRBs and the nature of dark matter.

Photons with energy half the axion mass can trigger the stimulated decay \cite{Kephart:1994uy,Rosa:2017ury,Caputo:2018vmy,Di:2023nnb,Dev:2023ijb} of axions, producing a echo. This phenomenon offers a novel opportunity to search for axion dark matter. By emitting a radio beam with high power into space and detecting the resulting echo, axion dark matter may be identified and studied \cite{Arza:2019nta,Arza:2021nec,Arza:2022dng,Arza:2023rcs,Gong:2023ilg}. In our previous work \cite{Di:2024snm}, we proposed that the stimulated decay of a dilute axion star associated with planet 9 could be triggered by directing a powerful radio beam at the star, resulting in an echo could be detected by current radio telescopes. Such an observation would distinguish an axion star from a PBH \cite{Scholtz:2019csj} or other planet 9 candidates. This method offers a promising avenue for exploring the axion parameter space constrained by FRBs.

This article is organized as follows. In Sec. II, we briefly introduce the concept of dilute axion stars. In Sec. III, we review the constraints on the axion parameter space imposed by FRBs. Section IV discusses the abundance of dilute axion stars in the Universe. In Sec. V, we analyze the potential for detecting echoes from dilute axion stars, as constrained by FRBs, within the Solar System by emitting powerful radio beams into these stars. Finally, Sec. VI presents our conclusions. In this article, we adopt natural units, setting $c = \hbar = 1$.

\section{Dilute Axion Stars}
The QCD axion is a hypothetical pseudo-Nambu-Goldstone boson with spin-0, characterized by a small mass $m_\phi$, weak self-interactions, and extremely weak couplings to Standard Model particles. The invariance of the axion Lagrangian under the shift symmetry $\phi(x) \rightarrow \phi(x) + 2\pi f_a$ requires the axion potential $V(\phi)$ to be periodic, such that $V(\phi) = V(\phi + 2\pi f_a)$, where $f_a$ represents the $U(1)$ symmetry breaking scale, also known as the axion decay constant. Performing a series expansion of the axion potential around $\phi = 0$, and considering only the first two leading terms, the potential can be expressed as
\bea
V(\phi)={1\over2} m_\phi^2 \phi^2+\frac{\lambda}{4!}\phi^4+{\cal O}\left(\frac{\lambda^2\phi^6}{6! m_\phi^2}\right),
\eea
where $\lambda = -\kappa m_\phi^2 / f_a^2$ is the coupling constant representing the attractive self-interaction of axions. The value of $\kappa$ depends on the choice of axion potential. For the instanton potential, $V(\phi) = (m_\phi f_a)^2 \left[1 - \cos\left({\phi}/{f_a}\right)\right]$, commonly used in axion phenomenology, $\kappa = 1$. For the chiral potential \cite{DiVecchia:1980yfw,GrillidiCortona:2015jxo}, $\kappa \simeq 0.34$ \cite{GrillidiCortona:2015jxo,Fujikura:2021omw}. Considering the interaction between axions and photons, the general Lagrangian for the axion is given by
\bea \label{axion}
{\cal L}&=&{1\over 2}\partial_{\mu}\phi\partial^{\mu}\phi-{1\over 2}m_{\phi}^2\phi^2
-\frac{\lambda}{4!}\phi^4 -{\cal O}\left(\frac{\lambda^2\phi^6}{6! m_\phi^2}\right)\nonumber\\
&&+\frac{\alpha K}{8\pi f_a}\phi F_{\mu\nu} \tilde F^{\mu\nu}-{1\over4} F_{\mu\nu}F^{\mu\nu},
\eea
where $\alpha$ is the fine-structure constant representing the electromagnetic field coupling strength, $K$ is a dimensionless quantity of order one that depends on the axion model, and $F_{\mu\nu}$ and $\tilde{F}^{\mu\nu}$ represent the electromagnetic field tensor and its dual, respectively. Due to their coupling with electromagnetic fields, axions can spontaneously decay into photons. This interaction has significant astrophysical implications and provides an experimental avenue for searching for these elusive particles. The spontaneous decay rate of an axion at rest into two photons is given by
\bea
\Gamma_\phi&=&\frac{\alpha^2 K^2 m_\phi^3}{256 \pi^3 f_a^2}\\
&=&1.02\times 10^{-50}~{\rm{s}^{-1}} K^2 \left( \frac{m_\phi}{10^{-5}~\rm{eV}} \right)^3
\left(\frac{f_a}{10^{12}~\rm{GeV}}\right)^{-2}.\nonumber
\eea
The extremely low decay rate makes it challenging to detect axions through photon signals from their spontaneous decay. For QCD axions, there exists a well-known relationship between the decay constant $f_a$ and the axion mass \cite{Weinberg:1977ma}:
\bea \label{relation}
m_{\phi}&=&\frac{\sqrt {m_u m_d}}{m_u+m_d}\frac{f_\pi m_\pi}{f_a}\nonumber\\
&&\simeq6\times10^{-6}~{\rm{eV}}\left(\frac{f_a}{10^{12}~\rm{GeV}}\right)^{-1},
\eea
where $m_u \simeq 2.2~\rm{MeV}$, $m_d \simeq 4.7~\rm{MeV}$, and $m_\pi \simeq 135~\rm{MeV}$ are the masses of the up quark, down quark, and pion, respectively, while $f_\pi \simeq 92~\rm{MeV}$ is the pion decay constant. This relationship is illustrated in Fig.~\ref{fig:fa}.
As bosons, axions can achieve exceptionally high phase space densities, potentially leading to the formation of BECs \cite{Sikivie:2009qn} and the creation of axion stars, which are divided into dilute and dense branches \cite{Schiappacasse:2017ham,Chavanis:2017loo,Visinelli:2017ooc,Eby:2019ntd}. Dilute axion stars are stable against perturbations, while dense axion stars may have lifetimes much shorter than any cosmological timescale \cite{Braaten:2019knj,Seidel:1991zh,Hertzberg:2010yz,Eby:2015hyx,Wang:2020zur}.

For a stable dilute axion star, equilibrium is maintained by the balance between attractive self-gravity and self-interactions, and the repulsive pressure from the Heisenberg uncertainty principle. This equilibrium persists as long as the star's density remains low, minimizing the influence of self-interactions. However, once the star's mass exceeds a critical threshold, determined by the strength of the attractive self-interaction coupling constant, the equilibrium is disrupted. The maximum mass $M_{\rm{max}}$ and corresponding minimum radius $R_{\rm{min}}$ of a stable axion star are given by \cite{Chavanis:2011zi,Chavanis:2011zm}
\bea
\label{maximum mass}
M_{\rm{max}}\sim 5.073\frac{M_{pl}}{\sqrt {|\lambda|}},~~~~~~R_{\rm{min}}\sim\sqrt {|\lambda|}\frac{M_{pl}}{m_\phi}\lambda_c,
\eea
where $\lambda_c$ is the Compton wavelength of the axion and $M_{pl}$ is the Planck mass.
When the mass of an axion star exceeds the critical mass given by Eq.~\eqref{maximum mass}, either due to merger events \cite{Mundim:2010hi,Cotner:2016aaq,Schwabe:2016rze,Eby:2017xaw,Hertzberg:2020dbk,Du:2023jxh,Maseizik:2024qly} or through the accretion of axions from the surrounding environment \cite{Chen:2020cef,Chan:2022bkz,Dmitriev:2023ipv}, self-interactions become significant, potentially leading to the collapse of the star \cite{Levkov:2016rkk,Chavanis:2016dab,Eby:2016cnq,Fox:2023xgx}.
Substituting the attractive coupling constant of self-interaction, $\lambda = -\kappa m_\phi^2 / f_a^2$, and the axion's Compton wavelength, $\lambda_c = 2\pi / m_\phi$, into Eq.~\eqref{maximum mass}, the critical mass and minimum radius are expressed as
\bea\label{critical mass}
M_{\rm{max}}&\sim&5.97\times10^{-12}M_{\odot}~\kappa^{-1/2}\nonumber\\
&&\times\left(\frac{m_\phi}{10^{-5}~{\rm{eV}}}\right)^{-1}
\left(\frac{f_a}{10^{12}~{\rm{GeV}}}\right),
\eea
\bea\label{radius}
R_{\rm{min}}&\sim&2.41\times10^2~{\rm{km}}~\kappa^{1/2}\nonumber\\
&&\times\left(\frac{m_\phi}{10^{-5}~{\rm{eV}}}\right)^{-1}
\left(\frac{f_a}{10^{12}~{\rm{GeV}}}\right)^{-1}.
\eea
These quantities depend not only on the mass and decay constant of the axion but also on the choice of axion potential. For a general dilute axion star, we denote its mass as $M_{\rm{AS}}$ and its radius as $R_{\rm{AS}}$.

\section{Constraints by fast radio bursts}
Despite the extremely low spontaneous decay rate, axion decay rates can be enhanced in the presence of ambient photons, as photons with energy equivalent to half of the axion mass can stimulate decays of axions into two photons. The change in photon number density $n_{\lambda^\prime}$ within a dilute axion star, resulting from axion decay and inverse decay, is governed by the Boltzmann equation \cite{Kephart:1994uy}:
\bea
\frac{dn_{\lambda^\prime}}{dt}&=&\int dX_{\rm{LIPS}}|{\cal M(\phi\rightarrow\gamma(\lambda^\prime)\gamma(\lambda^\prime))}|^2\nonumber\\
&&\times\{f_\phi(\textbf{p})[1+f_{\lambda^\prime}(\textbf{k}_1)][1+f_{\lambda^\prime}(\textbf{k}_2)]\nonumber\\
&&-f_{\lambda^\prime}(\textbf{k}_1)f_{\lambda^\prime}(\textbf{k}_2)[1+f_\phi(\textbf{p})]\},
\eea
where $\cal M(\phi \rightarrow \gamma(\lambda^\prime)\gamma(\lambda^\prime))$ refers to the matrix element corresponding to the coupling term between axions and photons, $f_\phi$ is the phase space density of axions, and $f_{\lambda^\prime}$ is the phase space density of photons with helicity $\lambda^\prime$.
After performing the phase space integration for the above equation and considering the two possible helicities of photons, the time evolution of the total photon number density becomes \cite{Kephart:1994uy, Rosa:2017ury}
\bea\label{photon number density}
\frac{dn_\gamma}{dt}=2\Gamma_\phi n_\phi+\frac{16\pi^2}{m_\phi^3v}\Gamma_\phi n_\phi n_\gamma
-\frac{16\pi^2}{3m_\phi^3}
\left(v+\frac{3}{2}\right)\Gamma_\phi n_\gamma^2,\nonumber\\
\eea
where $v$ is the maximum velocity of an axion within the axion star, which is roughly $1/(2R_{\rm{AS}}m_\phi)$ as estimated using the uncertainty principle.

\begin{figure}
\begin{center}
\includegraphics[width=0.45\textwidth]{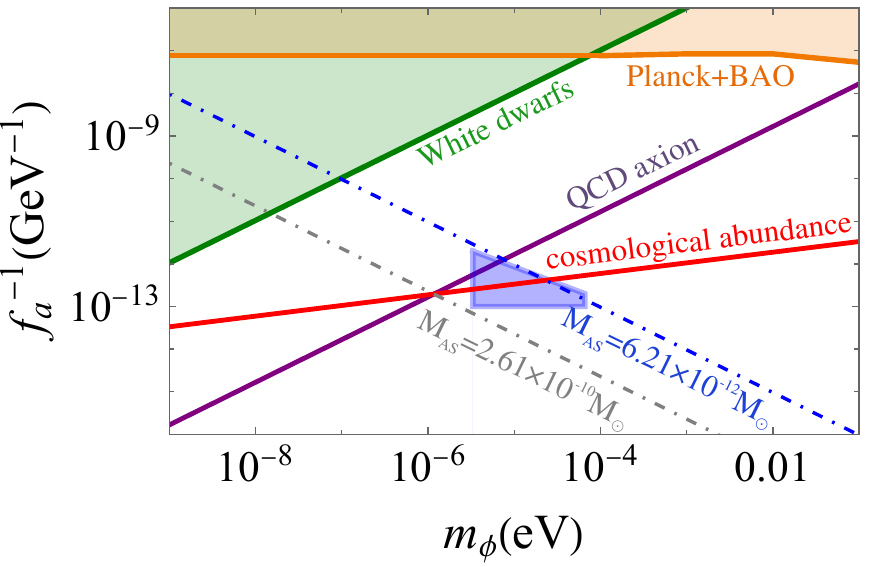}
\caption{Constraints on the axion parameter space. The region above the orange line is excluded based on Planck 2018 and BAO measurements \cite{Caloni:2022uya}. A recent study \cite{Springmann:2024ret} places a model-independent bound on QCD axions from supernova 1987A, which is stronger than the bounds from Planck and BAO. Constraints from white dwarf observations are represented by the green region \cite{Balkin:2022qer}, with recent updates provided in Ref. \cite{Gomez-Banon:2024oux}. The purple line indicates constraints for QCD axions, while the blue region corresponds to the parameter space constrained by FRBs, where the maximum mass of dilute axion stars ranges from $6.21\times10^{-12}M_\odot$ to $2.61\times10^{-10}M_\odot$.}
\label{fig:fa}
\end{center}
\end{figure}

In Eq. \eqref{photon number density}, the first term represents spontaneous axion decay, while the second term accounts for the stimulated decay of axions in the presence of ambient photons. The third term corresponds to the inverse decay process, in which photons are converted back into axions. The component with a factor of $3/2$ corresponds to the generation of ``sterile" axions with high velocity that escape from the axion star \cite{Kephart:1994uy}. Additionally, photons produced through spontaneous axion decay and stimulated decay in the presence of ambient photons can also escape from the axion star, with an escape rate of $\Gamma_e = 1/{R_{\rm{AS}}}$. Integrating Eq. \eqref{photon number density} and considering the photons that escape from the axion star, the coupled differential equations describing the evolution of the number of axions and photons within the axion star are given by \cite{Di:2024tlz}
\bea \label{coupled differential equation1}
\frac{dN_\gamma}{dt}&=&2\Gamma_\phi N_\phi+\frac{12\pi}{m_\phi^3vR_{\rm{AS}}^3}\Gamma_\phi N_\phi N_\gamma\nonumber\\
&&-\frac{2\pi (2v+3)}{m_\phi^3R_{\rm{AS}}^3}\Gamma_\phi N_\gamma^2-\Gamma_e N_\gamma,
\eea
\bea  \label{coupled differential equation2}
\frac{dN_\phi}{dt}&=&-\Gamma_\phi N_\phi-\frac{6\pi}{m_\phi^3vR_{\rm{AS}}^3}\Gamma_\phi N_\phi N_\gamma\nonumber\\
&&+\frac{4\pi v}{m_\phi^3R_{\rm{AS}}^3} \Gamma_\phi N_\gamma^2.
\eea
However, even for dilute axion stars in a critical state, the photon number $N_\gamma \simeq (2\Gamma_\phi/\Gamma_e)N_\phi$ resulting from spontaneous decay is insufficient to initiate stimulated decay.

When the axion star mass exceeds the critical threshold, it collapses. As the size approaches the critical radius $R_{cr}\simeq24\pi\Gamma_\phi M_{\rm{max}}/m_\phi^3$ \cite{Di:2024tlz}, stimulated decay initiates. The luminosity of the collapsing axion star from stimulated decay is given by \cite{Di:2024tlz}
\bea \label{luminosity}
L_\phi&=&\frac{1}{2}m_\phi N_\gamma R_{cr}^{-1}=\frac{N_\gamma m_\phi^4}{48\pi\Gamma_\phi M_{\rm{max}}}\nonumber\\
&\sim&3.60\times10^{40}{\rm{erg/s}}\left(\frac{N_\gamma}{10^{49}}\right)
\left(\frac{m_\phi}{10^{-5}{\rm{eV}}}\right)^2\left(\frac{f_a}{10^{12}{\rm{GeV}}}\right),\nonumber\\
\eea
where model-dependent constants $K=1$ and $\kappa=1$ are assumed. Detection of a radio signal consistent with the stimulated decay of a collapsing axion star could serve as a standard candle for constraining the Hubble constant, given the intense luminosity of such a radio burst \cite{Di:2024tlz}.

Interestingly, radio signals from the stimulated decay of collapsing axion stars may already have been detected. FRBs are bright, transient radio emissions lasting only milliseconds \cite{Lorimer:2007, Keane:2012yh, Thornton:2013iua}, with a frequency range spanning approximately 400 MHz to 8 GHz \cite{Petroff:2019tty} and total energy outputs typically between $10^{38}$ and $10^{40}$ erg \cite{Thornton:2013iua,Spitler:2014fla}. The photon frequency and total energy emitted by a collapsing axion star, for certain parameter ranges, align with those observed in FRBs \cite{Di:2024tlz}. For constraints on the axion parameter space derived from FRBs, see Ref. \cite{Di:2024tlz} and Fig.~\ref{fig:fa}. The constraints on the axion parameter space provide valuable guidance for experimental searches for axions. The maximum axion star mass within this constrained parameter space, assuming $\kappa=1$, ranges from approximately $6.21 \times 10^{-12} M_\odot$ to $2.61 \times 10^{-10} M_\odot$, as shown in Fig.~\ref{fig:fa}.

\section{Abundance of dilute axion stars}
The current relic density of axions is described by the equation \cite{Di:2023xaw}:
\bea \label{abundance}
\Omega_\phi h^2&\simeq&0.12
\left(\frac{g_\star(T_{\rm{osc}})}{106.75}\right)^{3/4}
\left(\frac{m_\phi}{10^{-6}~\rm{eV}}\right)^{1/2}\nonumber\\
&&\times\left(\frac{f_a}{5.32\times 10^{12}~\rm{GeV}}\right)^2.
\eea
This equation, applicable to axions produced via the misalignment mechanism, defines the lower bound of the axion parameter space, as shown in  Fig.~\ref{fig:fa}. By combining Eq. \eqref{relation} with Eq. \eqref{abundance}, we obtain the axion mass $m_\phi\simeq1.17 \times 10^{-6}$ eV and the decay constant $f_a\simeq5.11 \times 10^{12}$ GeV, assuming that QCD axions constitute the majority of dark matter. Substituting these values into Eq. \eqref{critical mass} and taking $\kappa=1$, we calculate the maximum mass of the corresponding dilute axion star to be approximately $2.61 \times 10^{-10} M_\odot$. This value coincides with the upper limit of the maximum axion star mass, as constrained by the parameter space derived from FRBs. As shown in Fig.~\ref{fig:fa}, the cosmological abundance line intersects this parameter space. This overlap significantly increases the likelihood of axions existing in this region and further supports the misalignment mechanism as a potential source of axions.

In the standard post-inflationary scenario, where $U(1)_{\rm{PQ}}$ symmetry is spontaneously broken after inflation, axion overdensities collapse during the radiation-dominated epoch, forming an early population of miniclusters with masses up to $10^{-12} M_\odot$ \cite{Hogan:1988mp,Kolb:1993zz,Kolb:1995bu}. By redshift $z=100$, about $75\%$ of axion dark matter resides within these bound structures \cite{Eggemeier:2019khm}, with minicluster masses reaching up to $10^{-9} M_\odot$ due to mergers. This closely matches the mass range of $6.21 \times 10^{-12} M_\odot$ to $2.61 \times 10^{-10} M_\odot$ for critical axion stars constrained by FRBs. Dilute axion stars evolve toward a critical state through the accretion of axions from the surrounding environment or via merger events. Once they surpass this critical threshold, they collapse, emitting strong radio signals \cite{Di:2023nnb} or relativistic axions \cite{Levkov:2016rkk}. After the collapse, a residual dilute axion star remains, retaining a mass close to its critical value and entering a subcritical state. This suggests that the likelihood of dilute axion stars being in either a critical or subcritical state is relatively high. For simplicity, we will assume that dilute axion stars are in a critical state in the following discussion.

The local dark matter density $\rho_{\rm{DM}}$ near the Solar System is estimated to be $0.01M_\odot~{\rm{pc}}^{-3}\approx0.4~\rm{GeV/cm^3}$, as inferred from stellar dynamics on scales greater than approximately 100 parsecs \cite{Read:2014qva,McMillan:2016jtx,Evans:2018bqy,deSalas:2020hbh}. Using this value, the local number density of dilute axion stars can be calculated as:
\bea \label{density}
n_{\rm{AS}}&=&\frac{\Omega_{\rm{AS}}}{\Omega_{\rm{DM}}}\frac{\rho_{\rm{DM}}}{M_{\rm{AS}}}\nonumber\\
&=& 1.27\times10^9~{\rm{pc}}^{-3}\left(\frac{ \Omega_{\rm{AS}}}{0.75\Omega_{\rm{DM}}}\right)
\left(\frac{6.21\times10^{-12} M_\odot}{M_{\rm{AS}}}\right),\nonumber\\
\eea
where $\Omega_{\rm{AS}}/{\Omega_{\rm{DM}}}$ represents the fraction of dilute axion stars within dark matter. Assuming $75\%$ of dark matter is composed of axion stars, a spherical volume with a radius of 1000 AU would contain approximately 605 axion stars with a mass of $6.21\times10^{-12}M_\odot$, or 14 axion stars with a mass of $2.61\times10^{-10}M_\odot$. The number density of dilute axion stars is inversely proportional to their mass: lower-mass axion stars are more numerous, increasing their likelihood of being present in the Solar System. This suggests that low-mass dilute axion stars, with maximum masses ranging from $6.21\times10^{-12}M_\odot$ to $2.61\times10^{-10}M_\odot$, could potentially be detected within the Solar System.
\section{Echo from Dilute Axion Stars}
A dilute axion star remains stable because the photon number resulting from spontaneous decay is insufficient to initiate stimulated decay. However, by transmitting a powerful radio beam with power $P$ to the axion star, with an angular frequency of photons equal to half the axion mass, the stimulated decay of axions can be significantly enhanced. As a result, the coupled differential equations given by Eq. \eqref{coupled differential equation1} and Eq. \eqref{coupled differential equation2} are modified as follows:
\bea
\frac{dN_\gamma}{dt}&=&2\Gamma_\phi N_\phi+\frac{12\pi}{m_\phi^3vR_{\rm{AS}}^3}\Gamma_\phi N_\phi (N_\gamma+N_{\gamma 0})\nonumber\\
&&-\frac{2\pi (2v+3)}{m_\phi^3R_{\rm{AS}}^3}\Gamma_\phi (N_\gamma+N_{\gamma 0})^2-\Gamma_e N_\gamma,
\eea
\bea
\frac{dN_\phi}{dt}&=&-\Gamma_\phi N_\phi-\frac{6\pi}{m_\phi^3vR_{\rm{AS}}^3}\Gamma_\phi N_\phi (N_\gamma+N_{\gamma 0})\nonumber\\
&&+\frac{4\pi v}{m_\phi^3R_{\rm{AS}}^3} \Gamma_\phi (N_\gamma+N_{\gamma 0})^2,
\eea
where $N_{\gamma 0}\simeq 2PR_{\rm{AS}}/m_\phi$ represents the number of photons from the radio beam transmitted into the axion star. By substituting Eq. \eqref{radius} into $N_{\gamma 0}$, we obtain
\bea
N_{\gamma 0}&\simeq&5.02\times10^{28}~\kappa^{1/2}\left(\frac{P}{50~\rm{MW}}\right)\nonumber\\
&&\times\left(\frac{m_\phi}{10^{-5}~\rm{eV}}\right)^{-2}
\left(\frac{f_a}{10^{12}~\rm{GeV}}\right)^{-1}.
\eea
Stimulated decay dominates over spontaneous decay when the condition ${6\pi}(N_\gamma+N_{\gamma 0})/({m_\phi^3vR_{\rm{AS}}^3})>1$ is satisfied. By transmitting a powerful radio beam with a power of $50~\rm{MW}$ into the axion star, we obtain $N_{\gamma 0}=3.86\times10^{26}$ and find that ${6\pi}N_{\gamma 0}/({m_\phi^3vR_{\rm{AS}}^3})=2.64\times10^{15}\gg1$, using the parameters $m_\phi=5\times10^{-5}$ eV, $f_a=5.20\times10^{12}$ GeV and taking the model-dependent constant $\kappa=1$. Thus, radio waves of sufficient power, such as $P=50~\rm{MW}$, can effectively induce the stimulated decay of a dilute axion star. The multi-beam relativistic klystron amplifier currently has the capability to output radio waves with a power of 1.047 GW \cite{sun:2023}, significantly exceeding the required 50 MW. This substantial power output makes it a promising candidate for emitting radio waves to interact with axion stars. In the following discussion, we will use the example of a 50 MW radio wave received by an axion star.
\begin{figure}
\begin{center}
\includegraphics[width=0.45\textwidth]{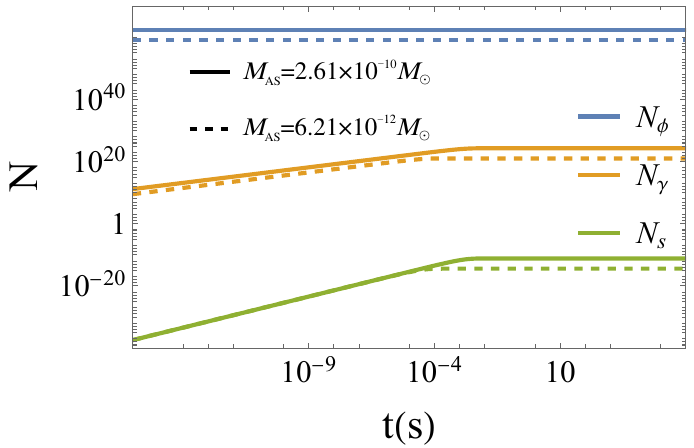}
\caption{Evolving axion, photon, and ``sterile'' axion numbers inside a dilute axion star under a $50{\rm{MW}}$ radio beam. The solid line represents an axion star with a mass of $M_{\rm{AS}}=2.61\times10^{-10}M_\odot$ and parameters $m_\phi=1.17\times10^{-6}~{\rm{eV}}$ and $f_a=5.11\times10^{12}~{\rm{GeV}}$. The dashed line corresponds to an axion star with a mass of $M_{\rm{AS}}=6.21\times10^{-12}M_\odot$ and parameters $m_\phi=5\times10^{-5}~{\rm{eV}}$ and $f_a=5.20\times10^{12}~{\rm{GeV}}$. Timing begins when the axion star is irradiated with radio beam. For $M_{\rm{AS}}=6.21\times10^{-12}M_\odot$, the stimulated decay of the dilute axion star stabilizes after approximately $10^{-4}$ seconds of irradiation, with a photon number of $N_\gamma=1.10\times10^{21}$.}
\label{fig:number}
\end{center}
\end{figure}

Figure~\ref{fig:number} presents two numerical solutions illustrating the evolution of the axion, photon, and ``sterile'' axion numbers for two representative parameter sets, assuming the model-dependent constants $K=1$ and $\kappa=1$. The solid line represents an axion star with a critical mass of $M_{\rm{AS}}=2.61\times10^{-10}M_\odot$ with parameters $m_\phi=1.17\times10^{-6}$ eV and a decay constant $f_a=5.11\times10^{12}$ GeV, representing the intersection of cosmological abundance and QCD axion constraints in Fig.~\ref{fig:fa}. The dashed line represents an axion star with a critical mass of $M_{\rm{AS}}=6.21\times10^{-12}M_\odot$, with parameters lying within the region constrained by FRBs: $m_\phi=5\times10^{-5}$ eV and $f_a=5.20\times10^{12}$ GeV. For the case $M_{\rm{AS}}=6.21\times10^{-12}~M_\odot$, the stimulated decay of the dilute axion star stabilizes after approximately $10^{-4}$ seconds of irradiation, resulting in a photon number $N_\gamma=1.10\times10^{21}$.
Considering the photons escaping from the dilute axion star, the luminosity can be expressed as:
\bea
L_\phi=\frac{1}{2} m_\phi N_\gamma\Gamma_e.
\eea
A notable feature of axion stimulated decay is that, in the axion rest frame, the emitted photons are restricted to propagate either in the same direction as or directly opposite to the incoming photons. However, within the dilute axion star, the velocity distribution of the axions leads to spatial dispersion of the emitted photons. Consequently, the luminosity flux observed from Earth is roughly given by
\bea
\label{flux}
F_\phi\sim{L_\phi\over 4\pi d^2}=\frac{m_\phi N_\gamma\Gamma_e}{8\pi d^2},
\eea
where $d$ denotes the distance between the dilute axion star and Earth. For a dilute axion star in its critical state, substituting Eq.~\eqref{radius} into $\Gamma_e=R_{\rm{AS}}^{-1}$, the photon escape rate becomes
\bea \label{gamma}
\Gamma_e\sim1.24\times10^3~{\rm{s^{-1}}}~\kappa^{-1/2}\left(\frac{m_\phi}{10^{-5}~{\rm{eV}}}\right)
\left(\frac{f_a}{10^{12}~{\rm{GeV}}}\right).\nonumber\\
\eea
By substituting Eq. \eqref{gamma} into Eq. \eqref{flux}, the flux can be expressed as
\bea
\label{flux2}
F_\phi&=&3.50\times10^{-35}~\kappa^{-1/2}\left(\frac{m_\phi}{10^{-5}~{\rm{eV}}}\right)^2\left(\frac{f_a}{10^{12}~{\rm{GeV}}}\right)\nonumber\\
&&\times\left(\frac{N_\gamma}{10^{20}}\right)\left(\frac{1000~\rm{AU}}{d}\right)^2{~\rm{W/cm^2}}.
\eea

\begin{figure}
\begin{center}
\includegraphics[width=0.45\textwidth]{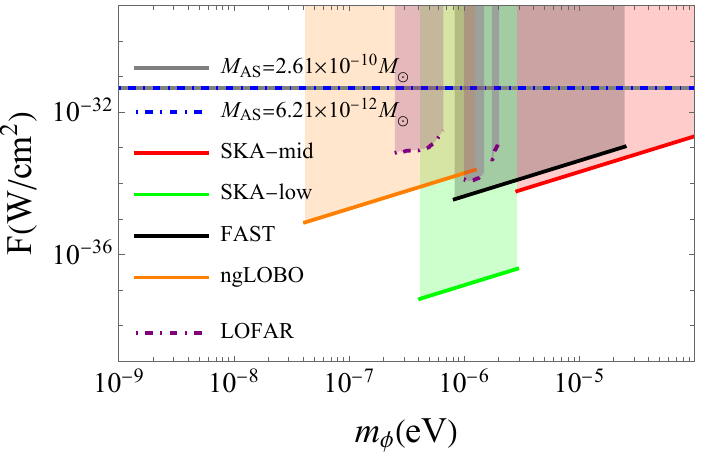}
\caption{Axion star flux at 1000 AU and the sensitivity of SKA, FAST, ngLOBO, and LOFAR for 1 hour observations. The sensitivity of these telescopes is detailed in Ref. \cite{Di:2024snm} and its references. Even for LOFAR, the flux of the echoes is one to two orders of magnitude higher than its sensitivity, making the signal easily detectable and distinguishable.}
\label{fig:flux}
\end{center}
\end{figure}

As illustrated in Fig.~\ref{fig:fa}, the critical mass of dilute axion stars constrained by FRBs ranges from approximately $6.21\times10^{-12}M_\odot$ to $2.61\times10^{-10}M_\odot$. Substituting Eq. \eqref{critical mass} into Eq. \eqref{flux2} for a mass of $6.21\times10^{-12}M_\odot$, the luminosity flux of the dilute axion star is calculated as
\bea
\label{flux3}
F_\phi&=&3.64\times10^{-34}\left(\frac{m_\phi}{10^{-5}~{\rm{eV}}}\right)^3\nonumber\\
&&\times\left(\frac{N_\gamma}{10^{21}}\right)\left(\frac{1000~\rm{AU}}{d}\right)^2~{\rm{W/cm^2}},
\eea
corresponding to the blue dot-dashed line with a flux of $F_\phi\approx 4.86\times10^{-32}~{\rm{W/cm^2}}$ for $d=1000$ AU, as shown in Fig.~\ref{fig:flux}.
Similarly, substituting Eq. \eqref{critical mass} into Eq. \eqref{flux2} for a mass of $2.61\times10^{-10}M_\odot$ gives
\bea
\label{flux3}
F_\phi&=&1.53\times10^{-32}\left(\frac{m_\phi}{10^{-6}~{\rm{eV}}}\right)^3\nonumber\\
&&\times\left(\frac{N_\gamma}{10^{24}}\right)\left(\frac{1000~\rm{AU}}{d}\right)^2~{\rm{W/cm^2}},
\eea
which corresponds to the gray line for $d=1000$ AU, effectively overlapping with the blue dot-dashed line, as shown in Fig.~\ref{fig:flux}.
Therefore, the luminosity flux of the echo from dilute axion stars at a distance of 1000 AU, as observed from Earth, is approximately $4.86\times10^{-32}~{\rm{W/cm^2}}$, corresponding to an axion star mass range of $6.21\times10^{-12}M_\odot$ to $2.61\times10^{-10}M_\odot$, as constrained by FRBs. This flux exceeds the sensitivity thresholds of radio telescopes such as SKA, FAST, ngLOBO, and LOFAR with an observation time of 1 hour, as shown in Fig.~\ref{fig:flux}. Thus, the decay signal is detectable by these radio telescopes.

Consequently, the stimulated decay of a dilute axion star within a volume of radius 1000 AU centered around Earth can be triggered by emitting a 50 MW radio beam into the star. The decay signal produces a spectral line that is almost monochromatic, with a frequency given by $f\simeq m_\phi/(4\pi) \approx 1.21(m_\phi/10^{-5}{\rm{eV}})~{\rm{GHz}}$, which serves as a distinctive feature of the signal. Detection of such an echo would confirm the existence of axion stars, solve the dark matter puzzle, and provide crucial evidence linking some FRBs to axion stars. Furthermore, FRBs originating from collapsing axion stars could serve as standard candles to help resolve the Hubble tension \cite{Di:2024tlz}.
If such echoes are not detected within a 1000 AU volume around Earth using this method, constraints can be placed on the fraction of dark matter composed of dilute axion stars with critical masses constrained by FRBs. This fraction could be roughly less than $5\%$ for a mass of $6.21\times10^{-12}M_\odot$ or $0.1\%$ for a mass of $2.61\times10^{-10}M_\odot$, depending on the axion parameters. This estimate assumes a uniform distribution of axion stars throughout the Universe.

\vspace{-5mm}
\section{Conclusions}
Axions are a promising candidate for dark matter. Due to their bosonic properties, they can reach exceptionally high phase space densities, which may lead to BEC and the formation of gravitationally bound objects known as axion stars. Collapsing axion stars with specific parameters may emit millisecond-long radio bursts with peak luminosities of approximately $10^{42}$ erg/s, which are consistent with the characteristics of observed non-repeating FRBs. If some FRBs are indeed produced by collapsing axion stars, this would impose constraints on the axion parameter space, offering valuable guidance for experimental axion searches. Thus, exploring the axion parameter space in relation to FRBs is crucial, as it relates both to the origin of FRBs and the nature of dark matter.

In the traditional post-inflationary scenario, axion stars could account for up to $75\%$ of dark matter, suggesting that some may exist within the Solar System. Building on previous work that explored the possibility of explaining planet 9 as a dilute axion star via stimulated decay induced by a 50 MW radio beam, we propose extending this method to search for dilute axion stars with masses ranging from $6.21\times10^{-12}M_\odot$ to $2.61\times10^{-10}M_\odot$. These masses are constrained by the characteristics of FRBs, which could potentially be explained by collapsing axion stars.
For axions within the parameter space constrained by FRBs, the echo resulting from the stimulated decay of axion stars could be detectable by ground-based radio telescopes. Detection of such an echo would not only confirm the existence of axion stars but also provide key evidence linking axion stars to FRBs. Furthermore, using FRBs from axion star collapses as standard candles could aid in resolving the Hubble tension. If no signal is detected, this method will provide constraints on the abundance of dark matter in the form of dilute axion stars with masses between $6.21\times10^{-12}M_\odot$ and $2.61\times10^{-10}M_\odot$. Under the assumption of a uniform distribution of dilute axion stars across the Universe, the abundance would be limited to less than approximately $5\%$ for masses of $6.21\times10^{-12}M_\odot$ or $0.1\%$ for masses of $2.61\times10^{-10}M_\odot$.

%%%%%%%%%%%%%%%%%%%%%%%%%%%%%%%%%%%%%%%%%%%%%%%%%%%%%%%%%%%%%%%%%%%%%%%%%%%%%
\begin{acknowledgments}
H. D. would like to thank Lijing Shao for useful discussions. This work was supported by National Natural Science Foundation of China under Grant No. 11947031 and East China University of Technology Research Foundation for Advanced Talents under Grant No. DHBK2019206.
\end{acknowledgments}
%%%%%%%%%%%%%%%%%%%%%%%%%%%%%%%%%%%%%%%%%%%%%%%%%%%%%%%%%%%%%%%%%%%%%%%%%%%%%


\begin{thebibliography}{999}

%\cite{Weinberg:1977ma}
\bibitem{Weinberg:1977ma}
S.~Weinberg,
A new light boson?,
\href{https://journals.aps.org/prl/abstract/10.1103/PhysRevLett.40.223}{Phys. Rev. Lett. \textbf{40}, 223 (1978)}.
%doi:10.1103/PhysRevLett.40.223
%5063 citations counted in INSPIRE as of 24 Jun 2023

%\cite{Wilczek:1977pj}
\bibitem{Wilczek:1977pj}
F.~Wilczek,
Problem of strong  $P$  and  $T$  invariance in the presence of instantons,
\href{https://journals.aps.org/prl/abstract/10.1103/PhysRevLett.40.279}{Phys. Rev. Lett. \textbf{40}, 279 (1978)}.
%doi:10.1103/PhysRevLett.40.279
%4862 citations counted in INSPIRE as of 24 Jun 2023

%\cite{Peccei:1977ur}
\bibitem{Peccei:1977ur}
R.~D.~Peccei and H.~R.~Quinn,
Constraints imposed by CP conservation in the presence of pseudoparticles,
\href{https://journals.aps.org/prd/abstract/10.1103/PhysRevD.16.1791}{Phys. Rev. D \textbf{16}, 1791 (1977)}.
%doi:10.1103/PhysRevD.16.1791
%3724 citations counted in INSPIRE as of 24 Jun 2023

%\cite{Peccei:1977hh}
\bibitem{Peccei:1977hh}
R.~D.~Peccei and H.~R.~Quinn,
CP conservation in the presence of pseudoparticles,
\href{https://journals.aps.org/prl/abstract/10.1103/PhysRevLett.38.1440}{Phys. Rev. Lett. \textbf{38}, 1440 (1977)}.
%doi:10.1103/PhysRevLett.38.1440
%7194 citations counted in INSPIRE as of 24 Jun 2023

%\cite{Arvanitaki:2009fg}
\bibitem{Arvanitaki:2009fg}
A.~Arvanitaki, S.~Dimopoulos, S.~Dubovsky, N.~Kaloper and J.~March-Russell,
String axiverse,
\href{https://doi.org/10.1103/PhysRevD.81.123530}{Phys. Rev. D \textbf{81}, 123530 (2010)}.
%doi:10.1103/PhysRevD.81.123530
%[arXiv:0905.4720 [hep-th]].
%1529 citations counted in INSPIRE as of 24 Jun 2023

%\cite{Preskill:1982cy}
\bibitem{Preskill:1982cy}
J.~Preskill, M.~B.~Wise and F.~Wilczek,
Cosmology of the invisible axion,
\href{https://www.sciencedirect.com/science/article/abs/pii/0370269383906378}{Phys. Lett. B \textbf{120}, 127 (1983)}.
%doi:10.1016/0370-2693(83)90637-8
%3037 citations counted in INSPIRE as of 24 Aug 2023

%\cite{Abbott:1982af}
\bibitem{Abbott:1982af}
L.~F.~Abbott and P.~Sikivie,
A cosmological Bbound on the invisible axion,
\href{https://www.sciencedirect.com/science/article/abs/pii/037026938390638X}{Phys. Lett. B \textbf{120}, 133 (1983)}.
%doi:10.1016/0370-2693(83)90638-X
%2807 citations counted in INSPIRE as of 24 Aug 2023

%\cite{Dine:1982ah}
\bibitem{Dine:1982ah}
M.~Dine and W.~Fischler,
The not so harmless axion,
\href{https://www.sciencedirect.com/science/article/abs/pii/0370269383906391}{Phys. Lett. B \textbf{120}, 137 (1983)}.
%doi:10.1016/0370-2693(83)90639-1
%2773 citations counted in INSPIRE as of 24 Aug 2023

%\cite{Co:2019jts}
\bibitem{Co:2019jts}
R.~T.~Co, L.~J.~Hall and K.~Harigaya,
Axion kinetic misalignment mechanism,
\href{https://doi.org/10.1103/PhysRevLett.124.251802}{Phys. Rev. Lett. \textbf{124}, 251802 (2020)}.
%doi:10.1103/PhysRevLett.124.251802
%[arXiv:1910.14152 [hep-ph]].
%128 citations counted in INSPIRE as of 14 Oct 2023

%\cite{Turner:1986tb}
\bibitem{Turner:1986tb}
M.~S.~Turner,
Early-universe thermal production of not-so-invisible axions,
\href{https://doi.org/10.1103/PhysRevLett.59.2489}{Phys. Rev. Lett. \textbf{59}, 2489 (1987);}
Erratum, \href{https://doi.org/10.1103/PhysRevLett.60.1101.3}{Phys. Rev. Lett. \textbf{60}, 1101 (1988)}.
%[erratum: Phys. Rev. Lett. \textbf{60}, 1101 (1988)].
%doi:10.1103/PhysRevLett.59.2489
%193 citations counted in INSPIRE as of 30 Nov 2024

%\cite{Salvio:2013iaa}
\bibitem{Salvio:2013iaa}
A.~Salvio, A.~Strumia and W.~Xue,
Thermal axion production,
\href{https://doi.org/10.1088/1475-7516/2014/01/011}{JCAP \textbf{01}, 011 (2014)}.
%doi:10.1088/1475-7516/2014/01/011
%[arXiv:1310.6982 [hep-ph]].
%181 citations counted in INSPIRE as of 30 Nov 2024

%\cite{Davis:1986xc}
\bibitem{Davis:1986xc}
R.~L.~Davis,
Cosmic axions from cosmic strings,
\href{https://doi.org/10.1016/0370-2693(86)90300-X}{Phys. Lett. B \textbf{180}, 225 (1986)}.
%doi:10.1016/0370-2693(86)90300-X
%465 citations counted in INSPIRE as of 04 Dec 2024

%\cite{Gorghetto:2020qws}
\bibitem{Gorghetto:2020qws}
M.~Gorghetto, E.~Hardy and G.~Villadoro,
More axions from strings,
\href{https://doi.org/10.21468/SciPostPhys.10.2.050}{SciPost Phys. \textbf{10}, 050 (2021)}.
%doi:10.21468/SciPostPhys.10.2.050
%[arXiv:2007.04990 [hep-ph]].
%164 citations counted in INSPIRE as of 14 Oct 2023

%\cite{Schiavone:2021imu}
\bibitem{Schiavone:2021imu}
F.~Schiavone, D.~Montanino, A.~Mirizzi and F.~Capozzi,
Axion-like particles from primordial black holes shining through the Universe,
\href{https://doi.org/10.1088/1475-7516/2021/08/063}{JCAP \textbf{08}, 063 (2021)}.
%doi:10.1088/1475-7516/2021/08/063
%[arXiv:2107.03420 [hep-ph]].
%33 citations counted in INSPIRE as of 30 Nov 2024

%\cite{Li:2022xqh}
\bibitem{Li:2022xqh}
T.~Li and R.~J.~Zhang,
Axionlike particles from primordial black hole evaporation and their detection in neutrino experiments,
\href{https://doi.org/10.1103/PhysRevD.106.095034}{Phys. Rev. D \textbf{106}, 095034 (2022)}.
%doi:10.1103/PhysRevD.106.095034
%[arXiv:2208.02696 [hep-ph]].
%7 citations counted in INSPIRE as of 30 Nov 2024

%\cite{Mazde:2022sdx}
\bibitem{Mazde:2022sdx}
K.~Mazde and L.~Visinelli,
The interplay between the dark matter axion and primordial black holes,
\href{https://doi.org/10.1088/1475-7516/2023/01/021}{JCAP \textbf{01}, 021 (2023)}.
%doi:10.1088/1475-7516/2023/01/021
%[arXiv:2209.14307 [astro-ph.CO]].
%28 citations counted in INSPIRE as of 30 Nov 2024

%\cite{Sikivie:2009qn}
\bibitem{Sikivie:2009qn}
P.~Sikivie and Q.~Yang,
Bose-Einstein condensation of dark matter axions,
\href{https://doi.org/10.1103/PhysRevLett.103.111301}{Phys. Rev. Lett. \textbf{103}, 111301 (2009)}.
%doi:10.1103/PhysRevLett.103.111301
%[arXiv:0901.1106 [hep-ph]].
%429 citations counted in INSPIRE as of 24 Jun 2023

%\cite{Braaten:2019knj}
\bibitem{Braaten:2019knj}
E.~Braaten and H.~Zhang,
Colloquium : The physics of axion stars,
\href{https://journals.aps.org/rmp/abstract/10.1103/RevModPhys.91.041002} {Rev. Mod. Phys. \textbf{91}, 041002 (2019)}.
%doi:10.1103/RevModPhys.91.041002
%51 citations counted in INSPIRE as of 24 Jun 2023

%\cite{Visinelli:2021uve}
\bibitem{Visinelli:2021uve}
L.~Visinelli,
Boson stars and oscillatons: A review,
\href{https://doi.org/10.1142/S0218271821300068}{Int. J. Mod. Phys. D \textbf{30}, 2130006 (2021)}.
%doi:10.1142/S0218271821300068
%[arXiv:2109.05481 [gr-qc]].
%43 citations counted in INSPIRE as of 24 Jun 2023

%\cite{Gorghetto:2024vnp}
\bibitem{Gorghetto:2024vnp}
M.~Gorghetto, E.~Hardy and G.~Villadoro,
More axion stars from strings,
\href{https://doi.org/10.1007/JHEP08(2024)126}{JHEP \textbf{08}, 126 (2024)}.
%doi:10.1007/JHEP08(2024)126
%[arXiv:2405.19389 [hep-ph]].
%7 citations counted in INSPIRE as of 16 Sep 2024

%\cite{Chang:2024fol}
\bibitem{Chang:2024fol}
J.~H.~Chang, P.~J.~Fox and H.~Xiao,
Axion stars: mass functions and constraints,
\href{https://doi.org/10.1088/1475-7516/2024/08/023}{JCAP \textbf{08}, 023 (2024)}.
%doi:10.1088/1475-7516/2024/08/023
%[arXiv:2406.09499 [hep-ph]].
%1 citations counted in INSPIRE as of 15 Sep 2024

%\cite{Zhang:2024bjo}
\bibitem{Zhang:2024bjo}
H.~Y.~Zhang,
Unified view of scalar and vector dark matter solitons,
\href{https://arxiv.org/abs/2406.05031}{arXiv:2406.05031}.
%[arXiv:2406.05031 [hep-ph]].
%6 citations counted in INSPIRE as of 17 Nov 2024

%\cite{Brown:2004yy}
\bibitem{Brown:2004yy}
M.~E.~Brown, C.~Trujillo and D.~Rabinowitz,
Discovery of a candidate inner Oort cloud planetoid,
\href{https://doi.org/10.1086/422095}{Astrophys. J. \textbf{617}, 645 (2004)}.
%doi:10.1086/424564
%[arXiv:astro-ph/0404456 [astro-ph]].
%36 citations counted in INSPIRE as of 23 Jun 2024

%\cite{Trujillo}
\bibitem{Trujillo}
C.~Trujillo, and S.~S.~Sheppard, A Sedna-like body with a perihelion of 80 astronomical units,
\href{https://www.nature.com/articles/nature13156}{Nature \textbf{507}, 471 (2014)}.

%\cite{Batygin:2016zsa}
\bibitem{Batygin:2016zsa}
K.~Batygin and M.~E.~Brown,
Evidence for a distant giant planet in the Solar System,
\href{https://doi.org/10.3847/0004-6256/151/2/22}{Astron. J. \textbf{151}, 22 (2016)}.
%doi:10.3847/0004-6256/151/2/22
%[arXiv:1601.05438 [astro-ph.EP]].
%57 citations counted in INSPIRE as of 23 Jun 2024

%\cite{Mroz:2017mvf}
\bibitem{Mroz:2017mvf}
P.~Mroz, A.~Udalski, J.~Skowron, R.~Poleski, S.~Kozlowski, M.~K.~Szymanski, I.~Soszynski, L.~Wyrzykowski, P.~Pietrukowicz, K.~Ulaczyk \textit{et al.},
No large population of unbound or wide-orbit Jupiter-mass planets,
\href{https://doi.org/10.1038/nature23276}{Nature \textbf{548}, 183 (2017)}.
%doi:10.1038/nature23276
%[arXiv:1707.07634 [astro-ph.EP]].
%96 citations counted in INSPIRE as of 23 Jun 2024

%\cite{Niikura:2019kqi}
\bibitem{Niikura:2019kqi}
H.~Niikura, M.~Takada, S.~Yokoyama, T.~Sumi and S.~Masaki,
Constraints on Earth-mass primordial black holes from OGLE 5-year microlensing events,
\href{https://doi.org/10.1103/PhysRevD.99.083503}{Phys. Rev. D \textbf{99}, 083503 (2019)}.
%doi:10.1103/PhysRevD.99.083503
%[arXiv:1901.07120 [astro-ph.CO]].
%217 citations counted in INSPIRE as of 24 Jun 2023

%\cite{Sugiyama:2021xqg}
\bibitem{Sugiyama:2021xqg}
S.~Sugiyama, M.~Takada and A.~Kusenko,
Possible evidence of axion stars in HSC and OGLE microlensing events,
\href{https://doi.org/10.1016/j.physletb.2023.137891}{Phys. Lett. B \textbf{840}, 137891 (2023)}.
%doi:10.1016/j.physletb.2023.137891
%[arXiv:2108.03063 [hep-ph]].
%8 citations counted in INSPIRE as of 26 Oct 2023

%\cite{Batygin-PR}
\bibitem{Batygin-PR}
K.~Batygin,  F.~C.~Adams, M.~E.~Brown, J.~C.~Becker, The planet nine hypothesis,
\href{https://doi.org/10.1016/j.physrep.2019.01.009}{Phys. Rep. \textbf{805}, 1 (2019)}.

%\cite{Di:2023xaw}
\bibitem{Di:2023xaw}
H.~Di and H.~Shi,
Can planet 9 be an axion star?,
\href{https://doi.org/10.1103/PhysRevD.108.103038}{Phys. Rev. D \textbf{108}, 103038 (2023)}.
%doi:10.1103/PhysRevD.108.103038
%[arXiv:2308.07263 [hep-ph]].
%1 citations counted in INSPIRE as of 21 Dec 2023

%\cite{Di:2023nnb}
\bibitem{Di:2023nnb}
H.~Di,
Stimulated decay of collapsing axion stars and fast radio bursts,
\href{https://doi.org/10.1140/epjc/s10052-024-12654-4}{Eur. Phys. J. C \textbf{84}, 283 (2024)}.
%doi:10.1140/epjc/s10052-024-12654-4
%[arXiv:2311.06860 [hep-ph]].
%2 citations counted in INSPIRE as of 23 Jun 2024

%\cite{Di:2024tlz}
\bibitem{Di:2024tlz}
H.~Di, L.~Shao, Z.~Yi and S.~B.~Kong,
Novel standard candle: Collapsing axion stars,
\href{https://doi.org/10.1103/PhysRevD.110.103031}{Phys. Rev. D \textbf{110}, 103031 (2024)}.
%[arXiv:2409.05120 [hep-ph]].
%0 citations counted in INSPIRE as of 07 Nov 2024

%\cite{Iwazaki:2014wka}
\bibitem{Iwazaki:2014wka}
A.~Iwazaki,
Axion stars and fast radio bursts,
\href{https://doi.org/10.1103/PhysRevD.91.023008}{Phys. Rev. D \textbf{91}, 023008 (2015)}.
%doi:10.1103/PhysRevD.91.023008
%[arXiv:1410.4323 [hep-ph]].
%73 citations counted in INSPIRE as of 12 Nov 2023

%\cite{Tkachev:2014dpa}
\bibitem{Tkachev:2014dpa}
I.~I.~Tkachev,
Fast radio bursts and axion miniclusters,
\href{https://doi.org/10.1134/S0021364015010154}{JETP Lett. \textbf{101}, 1 (2015)}.
%doi:10.1134/S0021364015010154
%[arXiv:1411.3900 [astro-ph.HE]].
%111 citations counted in INSPIRE as of 12 Nov 2023

%\cite{Raby:2016deh}
\bibitem{Raby:2016deh}
S.~Raby,
Axion star collisions with neutron stars and fast radio bursts,
\href{https://doi.org/10.1103/PhysRevD.94.103004}{Phys. Rev. D \textbf{94}, 103004 (2016)}.
%doi:10.1103/PhysRevD.94.103004
%[arXiv:1609.01694 [hep-ph]].
%38 citations counted in INSPIRE as of 12 Nov 2023

%\cite{Buckley:2020fmh}
\bibitem{Buckley:2020fmh}
J.~H.~Buckley, P.~S.~B.~Dev, F.~Ferrer and F.~P.~Huang,
Fast radio bursts from axion stars moving through pulsar magnetospheres,
\href{https://doi.org/10.1103/PhysRevD.103.043015}{Phys. Rev. D \textbf{103}, 043015 (2021)}.
%doi:10.1103/PhysRevD.103.043015
%[arXiv:2004.06486 [astro-ph.HE]].
%45 citations counted in INSPIRE as of 18 Nov 2023

%\cite{Kephart:1994uy}
\bibitem{Kephart:1994uy}
T.~W.~Kephart and T.~J.~Weiler,
Stimulated radiation from axion cluster evolution,
\href{https://journals.aps.org/prd/abstract/10.1103/PhysRevD.52.3226}{Phys. Rev. D \textbf{52}, 3226 (1995)}.
%doi:10.1103/PhysRevD.52.3226
%27 citations counted in INSPIRE as of 05 Oct 2023

%\cite{Rosa:2017ury}
\bibitem{Rosa:2017ury}
J.~G.~Rosa and T.~W.~Kephart,
Stimulated axion decay in superradiant clouds around primordial black holes,
\href{https://journals.aps.org/prl/abstract/10.1103/PhysRevLett.120.231102}{Phys. Rev. Lett. \textbf{120}, 231102 (2018)}.
%doi:10.1103/PhysRevLett.120.231102
%[arXiv:1709.06581 [gr-qc]].
%93 citations counted in INSPIRE as of 05 Oct 2023

%\cite{Caputo:2018vmy}
\bibitem{Caputo:2018vmy}
A.~Caputo, M.~Regis, M.~Taoso and S.~J.~Witte,
Detecting the stimulated decay of axions at radio frequencies,
\href{https://doi.org/10.1088/1475-7516/2019/03/027}{JCAP \textbf{03}, 027 (2019)}.
%doi:10.1088/1475-7516/2019/03/027
%[arXiv:1811.08436 [hep-ph]].
%87 citations counted in INSPIRE as of 18 Nov 2023

%\cite{Dev:2023ijb}
\bibitem{Dev:2023ijb}
P.~S.~B.~Dev, F.~Ferrer and T.~Okawa,
On the Galactic radio signal from stimulated decay of axion dark matter,
\href{https://doi.org/10.1088/1475-7516/2024/04/045}{JCAP \textbf{04}, 045 (2024)}.
%doi:10.1088/1475-7516/2024/04/045
%[arXiv:2311.13653 [hep-ph]].
%5 citations counted in INSPIRE as of 30 Nov 2024

%\cite{Arza:2019nta}
\bibitem{Arza:2019nta}
A.~Arza and P.~Sikivie,
Production and detection of an axion dark matter echo,
\href{https://doi.org/10.1103/PhysRevLett.123.131804}{Phys. Rev. Lett. \textbf{123}, 131804 (2019)}.
%doi:10.1103/PhysRevLett.123.131804
%[arXiv:1902.00114 [hep-ph]].
%46 citations counted in INSPIRE as of 23 Jun 2024

%\cite{Arza:2021nec}
\bibitem{Arza:2021nec}
A.~Arza and E.~Todarello,
Axion dark matter echo: A detailed analysis,
\href{https://doi.org/10.1103/PhysRevD.105.023023}{Phys. Rev. D \textbf{105}, 023023 (2022)}.
%doi:10.1103/PhysRevD.105.023023
%[arXiv:2108.00195 [hep-ph]].
%12 citations counted in INSPIRE as of 17 Jul 2024

%\cite{Arza:2022dng}
\bibitem{Arza:2022dng}
A.~Arza, A.~Kryemadhi and K.~Zioutas,
Searching for axion streams with the echo method,
\href{https://doi.org/10.1103/PhysRevD.108.083001}{Phys. Rev. D \textbf{108}, 083001 (2023)}.
%doi:10.1103/PhysRevD.108.083001
%[arXiv:2212.10905 [hep-ph]].
%10 citations counted in INSPIRE as of 17 Jul 2024

%\cite{Arza:2023rcs}
\bibitem{Arza:2023rcs}
A.~Arza, Q.~Guo, L.~Wu, Q.~Yang, X.~Yang, Q.~Yuan and B.~Zhu,
Listening for echo from the stimulated axion decay with the 21 centimeter array,
\href{https://doi.org/10.1016/j.scib.2024.08.003}{Sci. Bull. \textbf{69}, 2971 (2024)}.
%doi:10.1016/j.scib.2024.08.003
%[arXiv:2309.06857 [hep-ph]].
%7 citations counted in INSPIRE as of 03 Nov 2024

%\cite{Gong:2023ilg}
\bibitem{Gong:2023ilg}
Y.~Gong, X.~Liu, L.~Wu, Q.~Yang and B.~Zhu,
Detecting quadratically coupled ultralight dark matter with stimulated annihilation,
\href{https://doi.org/10.1103/PhysRevD.109.055026}{Phys. Rev. D \textbf{109}, 055026 (2024)}.
%doi:10.1103/PhysRevD.109.055026
%[arXiv:2308.08477 [hep-ph]].
%4 citations counted in INSPIRE as of 23 Jun 2024

%\cite{Di:2024snm}
\bibitem{Di:2024snm}
H.~Di, H.~Shi and Z.~Yi,
Detection of dilute axion stars with stimulated decay,
\href{https://doi.org/10.1103/PhysRevD.111.023011}{Phys. Rev. D \textbf{111}, 023011 (2025)}.
%[arXiv:2407.08436 [hep-ph]].
%0 citations counted in INSPIRE as of 13 Sep 2024

%\cite{Scholtz:2019csj}
\bibitem{Scholtz:2019csj}
J.~Scholtz and J.~Unwin,
What if planet 9 is a primordial black hole?,
\href{https://doi.org/10.1103/PhysRevLett.125.051103}{Phys. Rev. Lett. \textbf{125}, 051103 (2020)}.
%doi:10.1103/PhysRevLett.125.051103
%[arXiv:1909.11090 [hep-ph]].
%54 citations counted in INSPIRE as of 24 Jun 2023

%\cite{DiVecchia:1980yfw}
\bibitem{DiVecchia:1980yfw}
P.~Di Vecchia and G.~Veneziano,
Chiral dynamics in the large n limit,
\href{https://doi.org/10.1016/0550-3213(80)90370-3}{Nucl. Phys. B \textbf{171}, 253 (1980)}.
%doi:10.1016/0550-3213(80)90370-3
%805 citations counted in INSPIRE as of 28 Oct 2024

%\cite{GrillidiCortona:2015jxo}
\bibitem{GrillidiCortona:2015jxo}
G.~Grilli di Cortona, E.~Hardy, J.~Pardo Vega and G.~Villadoro,
The QCD axion, precisely,
\href{https://doi.org/10.1007/JHEP01(2016)034}{JHEP \textbf{01}, 034 (2016)}.
%doi:10.1007/JHEP01(2016)034
%[arXiv:1511.02867 [hep-ph]].
%734 citations counted in INSPIRE as of 04 Nov 2024

%\cite{Fujikura:2021omw}
\bibitem{Fujikura:2021omw}
K.~Fujikura, M.~P.~Hertzberg, E.~D.~Schiappacasse and M.~Yamaguchi,
Microlensing constraints on axion stars including finite lens and source size effects,
\href{https://doi.org/10.1103/PhysRevD.104.123012}{Phys. Rev. D \textbf{104}, 123012 (2021)}.
%doi:10.1103/PhysRevD.104.123012
%[arXiv:2109.04283 [hep-ph]].
%19 citations counted in INSPIRE as of 28 Oct 2024

%\cite{Schiappacasse:2017ham}
\bibitem{Schiappacasse:2017ham}
E.~D.~Schiappacasse and M.~P.~Hertzberg,
Analysis of dark matter axion clumps with spherical symmetry,
\href{https://doi.org/10.1088/1475-7516/2018/01/037}{JCAP \textbf{01}, 037 (2018);}
Erratum, \href{https://doi.org/10.1088/1475-7516/2018/03/E01}{JCAP \textbf{03}, E01 (2018)}.
%[erratum: JCAP \textbf{03}, E01 (2018)]
%doi:10.1088/1475-7516/2018/01/037
%[arXiv:1710.04729 [hep-ph]].
%109 citations counted in INSPIRE as of 06 Dec 2024

%\cite{Chavanis:2017loo}
\bibitem{Chavanis:2017loo}
P.~H.~Chavanis,
Phase transitions between dilute and dense axion stars,
\href{https://doi.org/10.1103/PhysRevD.98.023009}{Phys. Rev. D \textbf{98}, 023009 (2018)}.
%doi:10.1103/PhysRevD.98.023009
%[arXiv:1710.06268 [gr-qc]].
%90 citations counted in INSPIRE as of 25 Oct 2023

%\cite{Visinelli:2017ooc}
\bibitem{Visinelli:2017ooc}
L.~Visinelli, S.~Baum, J.~Redondo, K.~Freese and F.~Wilczek,
Dilute and dense axion stars,
\href{https://doi.org/10.1016/j.physletb.2017.12.010}{Phys. Lett. B \textbf{777}, 64 (2018)}.
%doi:10.1016/j.physletb.2017.12.010
%[arXiv:1710.08910 [astro-ph.CO]].
%165 citations counted in INSPIRE as of 25 Oct 2023

%\cite{Eby:2019ntd}
\bibitem{Eby:2019ntd}
J.~Eby, M.~Leembruggen, L.~Street, P.~Suranyi and L.~C.~R.~Wijewardhana,
Global view of QCD axion stars,
\href{https://doi.org/10.1103/PhysRevD.100.063002}{Phys. Rev. D \textbf{100}, 063002 (2019)}.
%doi:10.1103/PhysRevD.100.063002
%[arXiv:1905.00981 [hep-ph]].
%40 citations counted in INSPIRE as of 25 Oct 2023

%\cite{Seidel:1991zh}
\bibitem{Seidel:1991zh}
E.~Seidel and W.~M.~Suen,
Oscillating soliton stars,
\href{https://doi.org/10.1103/PhysRevLett.66.1659}{Phys. Rev. Lett. \textbf{66}, 1659 (1991)}.
%doi:10.1103/PhysRevLett.66.1659
%290 citations counted in INSPIRE as of 25 Oct 2023

%\cite{Hertzberg:2010yz}
\bibitem{Hertzberg:2010yz}
M.~P.~Hertzberg,
Quantum radiation of oscillons,
\href{https://doi.org/10.1103/PhysRevD.82.045022}{Phys. Rev. D \textbf{82}, 045022 (2010)}.
%doi:10.1103/PhysRevD.82.045022
%[arXiv:1003.3459 [hep-th]].
%119 citations counted in INSPIRE as of 25 Oct 2023

%\cite{Eby:2015hyx}
\bibitem{Eby:2015hyx}
J.~Eby, P.~Suranyi and L.~C.~R.~Wijewardhana,
The lifetime of axion stars,
\href{https://doi.org/10.1142/S0217732316500905}{Mod. Phys. Lett. A \textbf{31}, 1650090 (2016)}.
%doi:10.1142/S0217732316500905
%[arXiv:1512.01709 [hep-ph]].
%68 citations counted in INSPIRE as of 25 Oct 2023

%\cite{Wang:2020zur}
\bibitem{Wang:2020zur}
Z.~Wang, L.~Shao and L.~X.~Li,
Resonant instability of axionic dark matter clumps,
\href{https://doi.org/10.1088/1475-7516/2020/07/038}{JCAP \textbf{07}, 038 (2020)}.
%doi:10.1088/1475-7516/2020/07/038
%[arXiv:2002.09144 [hep-ph]].
%21 citations counted in INSPIRE as of 12 Oct 2024

%\cite{Chavanis:2011zi}
\bibitem{Chavanis:2011zi}
P.~H.~Chavanis,
Mass-radius relation of Newtonian self-gravitating Bose-Einstein condensates with short-range interactions: I. Analytical results,
\href{https://journals.aps.org/prd/abstract/10.1103/PhysRevD.84.043531}{Phys. Rev. D \textbf{84}, 043531 (2011)}.
%doi:10.1103/PhysRevD.84.043531
%[arXiv:1103.2050 [astro-ph.CO]].
%341 citations counted in INSPIRE as of 11 Aug 2023

%\cite{Chavanis:2011zm}
\bibitem{Chavanis:2011zm}
P.~H.~Chavanis and L.~Delfini,
Mass-radius relation of Newtonian self-gravitating Bose-Einstein condensates with short-range interactions: II. Numerical results,
\href{https://doi.org/10.1103/PhysRevD.84.043532}{Phys. Rev. D \textbf{84}, 043532 (2011)}.
%doi:10.1103/PhysRevD.84.043532
%[arXiv:1103.2054 [astro-ph.CO]].
%228 citations counted in INSPIRE as of 06 Oct 2023

%\cite{Mundim:2010hi}
\bibitem{Mundim:2010hi}
B.~C.~Mundim,
A numerical study of boson star binaries,
\href{https://arxiv.org/abs/1003.0239}{arXiv:1003.0239}.
%[arXiv:1003.0239 [gr-qc]].
%7 citations counted in INSPIRE as of 25 Oct 2023

%\cite{Cotner:2016aaq}
\bibitem{Cotner:2016aaq}
E.~Cotner,
Collisional interactions between self-interacting nonrelativistic boson stars: Effective potential analysis and numerical simulations,
\href{https://doi.org/10.1103/PhysRevD.94.063503}{Phys. Rev. D \textbf{94}, 063503 (2016)}.
%doi:10.1103/PhysRevD.94.063503
%[arXiv:1608.00547 [astro-ph.CO]].
%27 citations counted in INSPIRE as of 25 Oct 2023

%\cite{Schwabe:2016rze}
\bibitem{Schwabe:2016rze}
B.~Schwabe, J.~C.~Niemeyer and J.~F.~Engels,
Simulations of solitonic core mergers in ultralight axion dark matter cosmologies,
\href{https://doi.org/10.1103/PhysRevD.94.043513}{Phys. Rev. D \textbf{94}, 043513 (2016)}.
%doi:10.1103/PhysRevD.94.043513
%[arXiv:1606.05151 [astro-ph.CO]].
%226 citations counted in INSPIRE as of 25 Oct 2023

%\cite{Eby:2017xaw}
\bibitem{Eby:2017xaw}
J.~Eby, M.~Leembruggen, J.~Leeney, P.~Suranyi and L.~C.~R.~Wijewardhana,
Collisions of dark matter axion stars with astrophysical sources,
\href{https://doi.org/10.1007/JHEP04(2017)099}{JHEP \textbf{04}, 099 (2017)}.
%doi:10.1007/JHEP04(2017)099
%[arXiv:1701.01476 [astro-ph.CO]].
%37 citations counted in INSPIRE as of 25 Oct 2023

%\cite{Hertzberg:2020dbk}
\bibitem{Hertzberg:2020dbk}
M.~P.~Hertzberg, Y.~Li and E.~D.~Schiappacasse,
Merger of dark matter axion clumps and resonant photon emission,
\href{https://doi.org/10.1088/1475-7516/2020/07/067}{JCAP \textbf{07}, 067 (2020)}.
%doi:10.1088/1475-7516/2020/07/067
%[arXiv:2005.02405 [hep-ph]].
%47 citations counted in INSPIRE as of 25 Oct 2023

%\cite{Du:2023jxh}
\bibitem{Du:2023jxh}
X.~Du, D.~J.~E.~Marsh, M.~Escudero, A.~Benson, D.~Blas, C.~K.~Pooni and M.~Fairbairn,
Soliton merger rates and enhanced axion dark matter decay,
\href{https://doi.org/10.1103/PhysRevD.109.043019}{Phys. Rev. D \textbf{109}, 043019 (2024)}.
%doi:10.1103/PhysRevD.109.043019
%[arXiv:2301.09769 [astro-ph.CO]].
%23 citations counted in INSPIRE as of 23 Jun 2024

%\cite{Maseizik:2024qly}
\bibitem{Maseizik:2024qly}
D.~Maseizik and G.~Sigl,
Distributions and collision rates of ALP stars in the Milky~Way,
\href{https://doi.org/10.1103/PhysRevD.110.083015}{Phys. Rev. D \textbf{110}, 083015 (2024)}.
%doi:10.1103/PhysRevD.110.083015
%[arXiv:2404.07908 [astro-ph.CO]].
%8 citations counted in INSPIRE as of 11 Oct 2024

%\cite{Chen:2020cef}
\bibitem{Chen:2020cef}
J.~Chen, X.~Du, E.~W.~Lentz, D.~J.~E.~Marsh and J.~C.~Niemeyer,
New insights into the formation and growth of boson stars in dark matter halos,
\href{https://doi.org/10.1103/PhysRevD.104.083022}{Phys. Rev. D \textbf{104}, 083022 (2021)}.
%doi:10.1103/PhysRevD.104.083022
%[arXiv:2011.01333 [astro-ph.CO]].
%60 citations counted in INSPIRE as of 25 Oct 2023

%\cite{Chan:2022bkz}
\bibitem{Chan:2022bkz}
J.~H.~H.~Chan, S.~Sibiryakov and W.~Xue,
Condensation and evaporation of boson stars,
\href{https://doi.org/10.1007/JHEP01(2024)071}{JHEP \textbf{01}, 071 (2024)}.
%doi:10.1007/JHEP01(2024)071
%[arXiv:2207.04057 [astro-ph.CO]].
%24 citations counted in INSPIRE as of 23 Jun 2024

%\cite{Dmitriev:2023ipv}
\bibitem{Dmitriev:2023ipv}
A.~S.~Dmitriev, D.~G.~Levkov, A.~G.~Panin and I.~I.~Tkachev,
Self-similar growth of bose stars,
\href{https://doi.org/10.1103/PhysRevLett.132.091001}{Phys. Rev. Lett. \textbf{132}, 091001 (2024)}.
%doi:10.1103/PhysRevLett.132.091001
%[arXiv:2305.01005 [astro-ph.CO]].
%19 citations counted in INSPIRE as of 23 Jun 2024

%\cite{Levkov:2016rkk}
\bibitem{Levkov:2016rkk}
D.~G.~Levkov, A.~G.~Panin and I.~I.~Tkachev,
Relativistic axions from collapsing Bose stars,
\href{https://doi.org/10.1103/PhysRevLett.118.011301}{Phys. Rev. Lett. \textbf{118}, 011301 (2017)}.
%doi:10.1103/PhysRevLett.118.011301
%[arXiv:1609.03611 [astro-ph.CO]].
%122 citations counted in INSPIRE as of 25 Oct 2023

%\cite{Chavanis:2016dab}
\bibitem{Chavanis:2016dab}
P.~H.~Chavanis,
Collapse of a self-gravitating Bose-Einstein condensate with attractive self-interaction,
\href{https://doi.org/10.1103/PhysRevD.94.083007}{Phys. Rev. D \textbf{94}, 083007 (2016)}.
%doi:10.1103/PhysRevD.94.083007
%[arXiv:1604.05904 [astro-ph.CO]].
%98 citations counted in INSPIRE as of 16 Feb 2024

%\cite{Eby:2016cnq}
\bibitem{Eby:2016cnq}
J.~Eby, M.~Leembruggen, P.~Suranyi and L.~C.~R.~Wijewardhana,
Collapse of axion stars,
\href{https://doi.org/10.1007/JHEP12(2016)066}{JHEP \textbf{12}, 066 (2016)}.
%doi:10.1007/JHEP12(2016)066
%[arXiv:1608.06911 [astro-ph.CO]].
%97 citations counted in INSPIRE as of 26 Oct 2023

%\cite{Fox:2023xgx}
\bibitem{Fox:2023xgx}
P.~J.~Fox, N.~Weiner and H.~Xiao,
Recurrent axion stars collapse with dark radiation emission and their cosmological constraints,
\href{https://doi.org/10.1103/PhysRevD.108.095043}{Phys. Rev. D \textbf{108}, 095043 (2023)}.
%doi:10.1103/PhysRevD.108.095043
%[arXiv:2302.00685 [hep-ph]].
%23 citations counted in INSPIRE as of 23 Jun 2024

%\cite{Caloni:2022uya}
\bibitem{Caloni:2022uya}
L.~Caloni, M.~Gerbino, M.~Lattanzi and L.~Visinelli,
Novel cosmological bounds on thermally-produced axion-like particles,
\href{https://doi.org/10.1088/1475-7516/2022/09/021}{JCAP \textbf{09}, 021 (2022)}.
%doi:10.1088/1475-7516/2022/09/021
%[arXiv:2205.01637 [astro-ph.CO]].
%26 citations counted in INSPIRE as of 06 Jul 2024

%\cite{Springmann:2024ret}
\bibitem{Springmann:2024ret}
K.~Springmann, M.~Stadlbauer, S.~Stelzl and A.~Weiler,
A universal bound on QCD axions from supernovae,
\href{https://arxiv.org/abs/2410.19902}{arXiv:2410.19902}.
%[arXiv:2410.19902 [hep-ph]].
%1 citations counted in INSPIRE as of 06 Dec 2024

%\cite{Balkin:2022qer}
\bibitem{Balkin:2022qer}
R.~Balkin, J.~Serra, K.~Springmann, S.~Stelzl and A.~Weiler,
White dwarfs as a probe of exceptionally light QCD axions,
\href{https://doi.org/10.1103/PhysRevD.109.095032}{Phys. Rev. D \textbf{109}, 095032 (2024)}.
%doi:10.1103/PhysRevD.109.095032
%[arXiv:2211.02661 [hep-ph]].
%23 citations counted in INSPIRE as of 06 Jul 2024

%\cite{Gomez-Banon:2024oux}
\bibitem{Gomez-Banon:2024oux}
A.~G\'omez-Ba\~n\'on, K.~Bartnick, K.~Springmann and J.~A.~Pons,
Constraining light QCD axions with isolated neutron star cooling,
\href{https://doi.org/10.1103/PhysRevLett.133.251002}{Phys. Rev. Lett. \textbf{133}, 251002 (2024)}.
%doi:10.1103/PhysRevLett.133.251002
%[arXiv:2408.07740 [hep-ph]].
%8 citations counted in INSPIRE as of 08 Jan 2025

%\cite{Lorimer:2007}
\bibitem{Lorimer:2007}
D.~R.~Lorimer, M.~Bailes, M.~A.~Mclaughlin, D.~J.~Narkevic, and F.~Crawford,
A bright millisecond radio burst of extragalactic origin,
\href{https://www.science.org/doi/10.1126/science.1147532}{Science, \textbf{318}, 777 (2007)}.
%doi:10.1126/science.114753

%\cite{Keane:2012yh}
\bibitem{Keane:2012yh}
E.~F.~Keane, B.~W.~Stappers, M.~Kramer and A.~G.~Lyne,
On the origin of a highly-dispersed coherent radio burst,
\href{https://doi.org/10.1111/j.1745-3933.2012.01306.x}{Mon. Not. Roy. Astron. Soc. \textbf{425}, 71 (2012)}.
%doi:10.1111/j.1745-3933.2012.01306.x
%[arXiv:1206.4135 [astro-ph.SR]].
%94 citations counted in INSPIRE as of 23 Oct 2023

%\cite{Thornton:2013iua}
\bibitem{Thornton:2013iua}
D.~Thornton, B.~Stappers, M.~Bailes, B.~R.~Barsdell, S.~D.~Bates, N.~D.~R.~Bhat, M.~Burgay, S.~Burke-Spolaor, D.~J.~Champion, P.~Coster \textit{et al.},
A population of fast radio bursts at cosmological distances,
\href{https://doi.org/10.1126/science.1236789}{Science \textbf{341}, 53 (2013)}.
%doi:10.1126/science.1236789
%[arXiv:1307.1628 [astro-ph.HE]].
%747 citations counted in INSPIRE as of 23 Oct 2023

%\cite{Petroff:2019tty}
\bibitem{Petroff:2019tty}
E.~Petroff, J.~W.~T.~Hessels and D.~R.~Lorimer,
Fast radio bursts,
\href{https://doi.org/10.1007/s00159-019-0116-6}{Astron. Astrophys. Rev. \textbf{27}, 4 (2019)}.
%doi:10.1007/s00159-019-0116-6
%[arXiv:1904.07947 [astro-ph.HE]].
%372 citations counted in INSPIRE as of 23 Sep 2024

%\cite{Spitler:2014fla}
\bibitem{Spitler:2014fla}
L.~G.~Spitler, J.~M.~Cordes, J.~W.~T.~Hessels, D.~R.~Lorimer, M.~A.~McLaughlin, S.~Chatterjee, F.~Crawford, J.~S.~Deneva, V.~M.~Kaspi, R.~S.~Wharton \textit{et al.},
Fast radio burst discovered in the arecibo pulsar ALFA survey,
\href{https://doi.org/10.1088/0004-637X/790/2/101}{Astrophys. J. \textbf{790}, 101 (2014)}.
%doi:10.1088/0004-637X/790/2/101
%[arXiv:1404.2934 [astro-ph.HE]].
%415 citations counted in INSPIRE as of 12 Nov 2023

%\cite{Hogan:1988mp}
\bibitem{Hogan:1988mp}
C.~J.~Hogan and M.~J.~Rees,
Axion miniclusters,
\href{https://doi.org/10.1016/0370-2693(88)91655-3}{Phys. Lett. B \textbf{205}, 228 (1988)}.
%doi:10.1016/0370-2693(88)91655-3
%313 citations counted in INSPIRE as of 19 Sep 2024

%\cite{Kolb:1993zz}
\bibitem{Kolb:1993zz}
E.~W.~Kolb and I.~I.~Tkachev,
Axion miniclusters and bose stars,
\href{https://doi.org/10.1103/PhysRevLett.71.3051}{Phys. Rev. Lett. \textbf{71}, 3051 (1993)}.
%doi:10.1103/PhysRevLett.71.3051
%[arXiv:hep-ph/9303313 [hep-ph]].
%405 citations counted in INSPIRE as of 19 Sep 2024

%\cite{Kolb:1995bu}
\bibitem{Kolb:1995bu}
E.~W.~Kolb and I.~I.~Tkachev,
Femtolensing and picolensing by axion miniclusters,
\href{https://doi.org/10.1086/309962}{Astrophys. J. Lett. \textbf{460}, L25 (1996)}.
%doi:10.1086/309962
%[arXiv:astro-ph/9510043 [astro-ph]].
%158 citations counted in INSPIRE as of 19 Sep 2024

%\cite{Eggemeier:2019khm}
\bibitem{Eggemeier:2019khm}
B.~Eggemeier, J.~Redondo, K.~Dolag, J.~C.~Niemeyer and A.~Vaquero,
First simulations of axion minicluster halos,
\href{https://doi.org/10.1103/PhysRevLett.125.041301}{Phys. Rev. Lett. \textbf{125}, 041301 (2020)}.
%doi:10.1103/PhysRevLett.125.041301
%[arXiv:1911.09417 [astro-ph.CO]].
%91 citations counted in INSPIRE as of 19 Sep 2024

%\cite{Read:2014qva}
\bibitem{Read:2014qva}
J.~I.~Read,
The local dark matter density,
\href{https://doi.org/10.1088/0954-3899/41/6/063101}{J. Phys. G \textbf{41}, 063101 (2014)}.
%doi:10.1088/0954-3899/41/6/063101
%[arXiv:1404.1938 [astro-ph.GA]].
%537 citations counted in INSPIRE as of 19 Sep 2024

%\cite{McMillan:2016jtx}
\bibitem{McMillan:2016jtx}
P.~J.~McMillan,
The mass distribution and gravitational potential of the Milky Way,
\href{https://doi.org/10.1093/mnras/stw2759}{Mon. Not. Roy. Astron. Soc. \textbf{465}, 76 (2016)}.
%doi:10.1093/mnras/stw2759
%[arXiv:1608.00971 [astro-ph.GA]].
%361 citations counted in INSPIRE as of 14 Sep 2024

%\cite{Evans:2018bqy}
\bibitem{Evans:2018bqy}
N.~W.~Evans, C.~A.~J.~O'Hare and C.~McCabe,
Refinement of the standard halo model for dark matter searches in light of the Gaia Sausage,
\href{https://doi.org/10.1103/PhysRevD.99.023012}{Phys. Rev. D \textbf{99}, 023012 (2019)}.
%doi:10.1103/PhysRevD.99.023012
%[arXiv:1810.11468 [astro-ph.GA]].
%206 citations counted in INSPIRE as of 19 Sep 2024

%\cite{deSalas:2020hbh}
\bibitem{deSalas:2020hbh}
P.~F.~de Salas and A.~Widmark,
Dark matter local density determination: recent observations and future prospects,
\href{https://doi.org/10.1088/1361-6633/ac24e7}{Rept. Prog. Phys. \textbf{84}, 104901 (2021)}.
%doi:10.1088/1361-6633/ac24e7
%[arXiv:2012.11477 [astro-ph.GA]].
%122 citations counted in INSPIRE as of 19 Sep 2024

%\cite{sun:2023}
\bibitem{sun:2023}
L.~Sun et al.,
Compact, stable, repetitive GW-level S-band multibeam relativistic klystron amplifier operating over a longer period in a lower magnetic field,
\href{https://doi.org/10.1109/TED.2023.3325415}{IEEE Trans. Electron Devices \textbf{70}, 6571 (2023)}.

\end{thebibliography}
\end{document}